\documentclass[twocolumn,showpacs,preprintnumbers,amsmath,amssymb]{revtex4}

\usepackage{graphicx}
\usepackage{dcolumn}
\usepackage{bm}

\begin{document}

\preprint{APS/123-QED}

\title{Comment on ''Direct space-time observation of pulse tunneling in an electromagnetic band gap''}

\author{G. Nimtz}
 \altaffiliation [Also at ]{II.Physikalisches Institut, Universit\"at zu K\"oln}
\author{A.A. Stahlhofen}%
 \email{G.Nimtz@uni-koeln.de}
\affiliation{Institut f\"ur Integrierte Naturwissenschaft,
Universit\"at Koblenz\\Universi\"atsstrasse 1, 56070 Koblenz,
Germany}


\date{\today}

\begin{abstract}
The investigation presented by Doiron, Hache, and Winful [Phys.
Rev. A 76, 023823 (2007)] is not valid for the tunneling process
as claimed in the paper.
\end{abstract}

\maketitle

The title of a recent article by Doiron et al.~\cite{Winful} is
misleading. The authors have investigated a dielectric mirror but
not a tunneling barrier, see also Refs.\cite{Winful2}. The
measured and discussed superluminal group velocity  is similar to
that studied on a Lorentz-Lorenz oscillator by Sommerfeld and
Brillouin a hundred years ago~\cite{Brillouin}. A dielectric
mirror is based on a periodical quarter wavelength structure of
two refractive indices in which destructive interference causes a
standing wave pattern decaying exponentially with mirror thickness
as shown in Figs. 2(a - f) in Ref.\cite{Winful}. There are nodes
in the intensity every half wavelength.

Tunneling, however, is understood and performed by electromagnetic
evanescent modes or by tunneling solutions of the Schr\"odinger
equation, which have purely imaginary wave numbers. The latter
includes a purely imaginary refractive index. Signals with purely
evanescent frequency components may travel at a superluminal
velocity ~\cite{NimtzH,Nimtz1}. Inside the barrier there is no
standing wave and tunneling proceeds even instantaneously, it
represents a process well described by virtual
photons~\cite{Stahlhofen}.

Recently, the transmission of dielectric mirrors and equivalent
structures has often been misinterpreted as tunneling, see for
instance Refs.~\cite{Steinberg,NimtzH}.

Actually, in the paper there are some errors: Fig.3 shows the
vacuum light velocity and in section (II,D) the dwell time is not
directly measured  but it is derived from an approximately
integrated stored energy and from the measured input power. This
indirect measuring procedure is indirect like the light
measurement from direct frequency and wavelength measurements, see
for information NBS and NIST, for instance.

In addition it is claimed to have measured a resonator decay time,
but detectors measure the traversal time of a pulse through a
black box independent of the content of the box. The authors are
asking \emph{whether an identifiable pulse peak actually
propagates through the barrier?} According to their Fig.1 not only
the peak but also the pulse half width (for instance representing
a digital signal) propagated faster than light and were correctly
detected. Figure 2(a-f) displays  the standing wave patterns at
various times. Figure 2(d) shows the peak of the pulse from the
mirrors entrance to the end of the mirror. Taking the x = 0 values
from Figs.2(a) to 2(f) we see the peak in 2(d) instantaneously
propagating from the entrance to the exit with the exponential
decay of the standing wave.

Describing the traversal time of a mirror as a decay time of a
cavity does not represent a contribution to the understanding of
the tunneling time as claimed by Doiron et al.~\cite{Winful}.


\end{document}